\documentclass[twocolumn]{aa}
\usepackage{natbib}
\bibpunct{(}{)}{;}{a}{}{,}
\usepackage{psfig}
\usepackage{epsfig,afterpage} 
\usepackage{graphicx}
\usepackage{pifont}
\usepackage{amssymb}
\usepackage{longtable}
\usepackage{array}
\usepackage{amsmath}
\usepackage{rotating}
\usepackage{multirow}
\usepackage{lscape}

\begin{document}

\title { An investigation of chromospheric
  activity spanning  the Vaughan--Preston  gap: impact on stellar ages.  \thanks{
    Based on  observations collected at the  European Organisation for
    Astronomical  Research in the  Southern Hemisphere,  Chile, during
    the observing run 073.D-0655.}}  \subtitle{ }

\author{ G. Pace \inst{1},  J. Melendez \inst{1}, L. Pasquini\inst{2},
  G.   Carraro  \inst{3}, J.   Danziger  \inst{4,5},  P.  Fran\c  cois
  \inst{6}, F. Matteucci\inst{4,5}, \and N. C. Santos \inst{1}}

\offprints{G. Pace, \email gpace@astro.up.pt}

\institute{
            Centro de Astrofisica, Universidade do Porto, 
            Rua das Estrelas, 4150--762 Porto, Portugal     
            \and
            European Southern Observatory, Karl Schwarzschildstr. 2,
            Garching bei M\"{u}nchen, Germany 
            \and
            European Southern Observatory, Casilla 19001, Santiago, Chile
            \and
            INAF, Osservatorio Astronomico di Trieste, 
            via G.B. Tiepolo 11, 34131 Trieste, Italy
            \and
            Department of Astronomy, University of Trieste, 
            via G.B. Tiepolo 11, 34131 Trieste, Italy
            \and
            Observatoire de  Paris, 64  Avenue de l'Observatoire, 75014
            Paris, France}


          \abstract {Chromospheric  activity is widely used  as an age
            indicator  for  solar--type   stars  based  on  the  early
            evidence that  there is a smooth evolution  from young and
            active   to  old  and   inactive  stars.    However,  this
            transition  may   require  modification  as  chromospheric
            activity  is not a  viable age  indicator for  stars older
            than  1 Gyr.}{We analysed  chromospheric activity  in five
            solar--type stars in two  open clusters, in order to study
            how chromospheric activity evolves  with time.}  { We took
            UVES high--resolution,  high S/N ratio spectra  of 3 stars
            in IC~4756 and  2 in NGC~5822, which were  combined with a
            previously  studied  data-set  and reanalysed  here.   The
            emission core of  the deep, photospheric Ca II  K line was
            used as  a probe of the chromospheric  activity.}  {All of
            the 5 stars in the  new sample, including those in the 1.2
            Gyr--old  NGC~5822,  have  activity levels  comparable  to
            those of Hyades and Praesepe.  }  {A likely interpretation
            of  our  data   is  that  solar--type--star  chromospheric
            activity, from  the age  of the Hyades  until that  of the
            Sun, does  not evolve smoothly.  Stars  change from active
            to inactive, crossing  the activity range corresponding to
            the Vaughan--Preston  gap, on a time--scale  that might be
            as short  as 200 Myr.   Evolution before and after  such a
            transition  is  much less  significant  than cyclical  and
            long--term variations.  We show that data presented in the
            literature  to  support  a  correlation  between  age  and
            activity  could  be also  interpreted  differently in  the
            light  of our  results.  Suggestions  have  been published
            that  relevant  stellar   structures  and/or  dynamos  are
            different for active and  inactive stars.  These provide a
            natural explanation  for the observations  presented here.
            More observations are required  in order to strengthen our
            results,  especially a  long--term  follow up  of our  two
            targets in the 1.2--Gyr old cluster NGC~5822.}


\authorrunning{Pace et al.\ }

\titlerunning{Chromospheric Activity.}

\maketitle

\section{Introduction}

It was already noticed in the early 60s \citep{W63} that chromospheric
activity   decreases during the main  sequence  life time of late type
stars like the Sun, and that the strength  of the emission in the core
of  the  broad and deep  Ca {\sc   II} H  and K   lines is  a reliable
indicator of chromospheric activity   level.  From that time   on, the
possibility of using H and K emission to measure stellar ages has been
widely explored    and  an impressive   amount    of data   on stellar
chromospheric  activity has been  collected,  especially in the  Mount
Wilson   program, started by  \cite{MWC},   and, more recently, during
planet--search surveys \citep[see   e.g.][]   {pss}.    Unfortunately,
chromospheric activity as an age indicator suffers a major limitation.
Activity   cycles equivalent  to    the   11--year long   solar   one,
longer--term activity variations, which caused the Maunder--minimum in
the Sun  , and, in some  stars, a  hidden  dependence on  stellar mass
\citep{cc07} and rotational  modulation, strongly affect  the level of
activity.  We believe that there is  an even more severe limitation on
the use of  chromospheric  activity  as an  age indicator:  its  decay
almost stops after about 1.5 Gyr, therefore activity is not a reliable
indicator of age for stars older than that.

Two  papers support  this view:  \cite{paper1}, hereafter  paper~1, of
which   this   study   can   be   considered   a   continuation,   and
\cite{lyra}. Nevertheless, many other  works pointed out the existence
of  a correlation  between chromospheric  activity and  age,  and many
attempts    were     made    to    calibrate     chromospheric    ages
\citep{sod91,don93,lach99}.  It is of primary importance to understand
the limitations of chromospheric activity as a reliable age indicator,
since it  has been  used to study,  for example,  the age--metallicity
relation  in the  Galactic disk  \citep{rp00a}, and  the age  of stars
hosting planetary systems \citep{saffe}.

In order to understand how chromospheric activity decays, observations
in solar--type stars  in open clusters are the  most appropriate tool,
since their ages can be  determined much more precisely than for field
stars.  The  paucity of old  open clusters in the  solar neighbourhood
hindered our progress until the  advent of the 8-- and 10--meter class
telescopes,  which  made  possible   the  acquisition  of  spectra  of
solar--type stars in several old  and distant clusters at a resolution
and a signal-to-noise ratio high enough to distinguish and resolve the
emission  core  in  the Ca{\sc  II}  K  line.   This is  necessary  to
accurately correct for the absorption by interstellar lines of Ca {\sc
  II} and thus to reliably measure chromospheric activity.

To the sample analysed in paper~1, we have added data on 5 solar--type
stars in the open clusters  NGC~5822 and IC~4756.  We selected the two
new target  clusters mainly  on the basis  of their  age, specifically
because we  wanted their ages to  lie between those of  the Hyades and
Praesepe and those of IC~4651 and NGC~3680, studied by \cite{sal}.  As
a result we have chromospheric activity data for 25 stars younger than
1 Gyr,  2 stars at  about 1 Gyr,  7 stars at about  1.5 Gyr, and  6 at
solar age.

The availability of UVES spectra allowed us to also accurately measure
temperatures and  metallicities for our target stars,  which have been
used in the  present work, but the details of  which will be published
elsewhere.

\section{Observations and sample}
\label{obs}

Target stars  were selected to be high--probability  single members of
IC~4756 and NGC~5822 and to be as similar to the Sun as possible.  The
choice  was made  on  the basis  of  the photometry  by \cite{hs}  for
IC~4756 and  by \cite{tatm} for  NGC~5822.  The former study  was made
with  photographic  plates,  and  the cluster  suffers  from  variable
extinction, therefore the photometric precision was not good enough to
exclude binaries with a high  level of confidence.  On the other hand,
probabilities of  membership published therein  are based on  a proper
motion study  which uses  a time baseline  of 43 years,  therefore are
quite accurate.  Spectra were taken  with the UVES spectrograph at the
focus of  the Kuyen telescope.  Observations were  carried out between
April and  September 2004.   The wavelength coverage  is from  3200 to
4600 {\AA} for the  blue arm, and from 4800 to 6800  {\AA} for the red
arm.  The slit width for the former was set to 0.8 arcsec, and for the
latter  to 0.4  arcsec,  giving a  resolution  of R$\approx$60000  and
R$\approx$100000 respectively  \citep{UVES}.  Data were  reduced using
the UVES pipeline \citep{UVESpipe}.

From  2 to  7 spectra  were obtained  for each  star.  Radial velocity
measurements  were   used  to  strengthen   single--member  selection.
Details   will   be   given   elsewhere   (Pace  et   al.   2009,   in
preparation). The final new sample consists of 3 stars in IC~4756 and 2 in
NGC~5822.

To the best  of our knowledge, \cite{sal} provide  the most recent set
of age evaluations for a sample  of open clusters that includes all of
those analysed by us.  They find an age of 0.7~Gyr for both Hyades and
Praesepe,  0.8 Gyr  for IC~4756,  1.2 Gyr  for NGC~5822,  1.4  Gyr for
NGC~3680, 1.7 Gyr for IC~4651, and 4.3 Gyr for M~67.

\section{Data analysis and results}

\label{CAda}

As an indicator of chromospheric activity we used $F^{\prime}_K$, i.e.
the energy flux of the Ca  {\sc II} K~line emitted per unit surface in
the  chromosphere.   It  is  common  practice  in  the  literature  to
normalise chromospheric  fluxes to bolometric emission and  to use the
logarithmic scale \citep{noyes}.  Hence we also used as a proxy of the
chromospheric activity:

$\log   R^{\prime}_K  =  \log   \left(   {F^{\prime}_K} / {\sigma
    T_{eff}^4} \right)$.

Normalisation of the  spectra was done as in  paper~1. From paper~1 we
also  took  1--{\AA}~K~index measurements  for  NGC~3680, IC~4651  and
M~67.  Index  measurements in Praesepe and Hyades  stars were slightly
revised  with respect  to the  values  published in  paper~1: the  new
indices are about 15 m{\AA} higher.  The stars in IC~4756 and NGC~5822
showed the  redshifted interstellar  K absorption line  which affected
the chromospheric K--line feature.  To measure the 1--{\AA}~K~index in
these  stars,   we  integrated  the   normalised  flux  only   in  the
uncontaminated  part  of  their   profile.   The  Praesepe  stars  are
unaffected by  IS absorption,  thus allowing us  to calculate  a ratio
between the flux measured over the 1--{\AA} and that measured over the
portion of  the feature which  is unaffected by interstellar  lines in
all  stars.  The  initial measures  for the  affected stars  were then
multiplied by this factor  to give a final corrected 1--{\AA}~K~index.
The errors  involved in the  measurement of the  1--{\AA}~K~index were
evaluated to  be within  6\%.  The 1--{\AA}~K~index  and chromospheric
activity  measurements, along  with stellar  parameters, are  given in
Table \ref{table1}.

\begin{figure}
\begin{center}
\includegraphics[width=8cm]{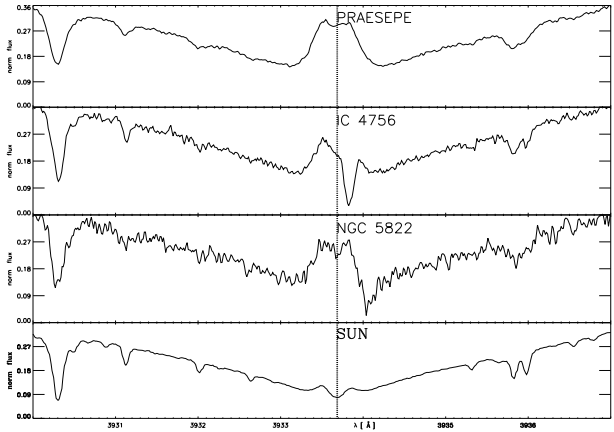}
\caption{The mean cluster Ca {\sc  II} K--line emission of IC~4756 and
  NGC~5822 is  compared with that of  Praesepe and with  the Sun.  All
  the  spectra are  normalised in  the same  way. Note  the  narrow IS
  absorption lines in the spectra of IC 4756 and NGC 5822.}
\label{figsp} 
\end{center}
\end{figure}

In Figure \ref{figsp} we show  the average cluster spectra for IC~4756
and  NGC~5822 and,  for  comparison,  that of  Praesepe  and the  solar
one. The interstellar feature is apparent in IC~4756 and NGC~5822.

For  the  Sun,  we  used  the 1--{\AA}~K~index  measurements  made  by
\cite{wl81}.   They monitored  solar chromospheric  activity  from the
first minimum to the maximum of the 21$^{st}$ solar--activity cycle.

Subtraction of  the photospheric contribution  to the 1--{\AA}~K~index
was made  as follows.  For the  Sun, we adopted  the same photospheric
correction  as   in  paper~1,  computed   using  a  solar--photosphere
synthetic spectrum  (courtesy of P. Bonifacio).  For  the other stars,
the  photospheric  contribution  was  computed by  scaling  the  solar
photospheric  contribution by  a factor  that depends  on  the stellar
parameters.  These  scaling factors  were computed using  the spectral
synthesis code of MOOG \citep[][version 2002]{MOOG}, and Kurucz's grid
of  models  \citep{kur93}.   Stellar parameters,  namely  temperature,
gravity,   microturbulence   and    metallicity,   were   known   from
\cite{paulson}  for the  Hyades stars,and  from \cite{papoc1}  for the
remainder of  the old sample, and  from our chemical  analysis for the
new sample (Pace et al. 2009, in preparation).

In  order to transform  the 1--{\AA}~K~index,  which is  an equivalent
width, into an intrinsic flux , namely $F^{\prime}_K$, we proceeded as
in paper~1, except  that we did not use  published photometry, but we
transformed  stellar temperatures and  metallicities into  B-V colours
inverting  equation  1)  in  \cite{sergio}.  For the  Sun  we  adopted
B-V=0.65~mag \citep{lcbiaz}.  The  use of stellar temperatures instead
of published  colours avoids the  cluster--to--cluster bias introduced
by the error in the cluster reddening.

Because of the  way we measured the index in the  new sample of stars,
we preferred not to use a  triangular filter in the integration of the
index, as  done, for instance,  in \cite{pauls02}.  Owing also  to the
lack of  common standards, we  could not make our  $\log R^{\prime}_K$
coincident with the widely used chromospheric activity indicator $\log
R^\prime_{HK}$.

\section{Discussion.}

\begin{figure*}
\begin{center}
\begin{tabular}{cc}

\includegraphics[width=8cm]{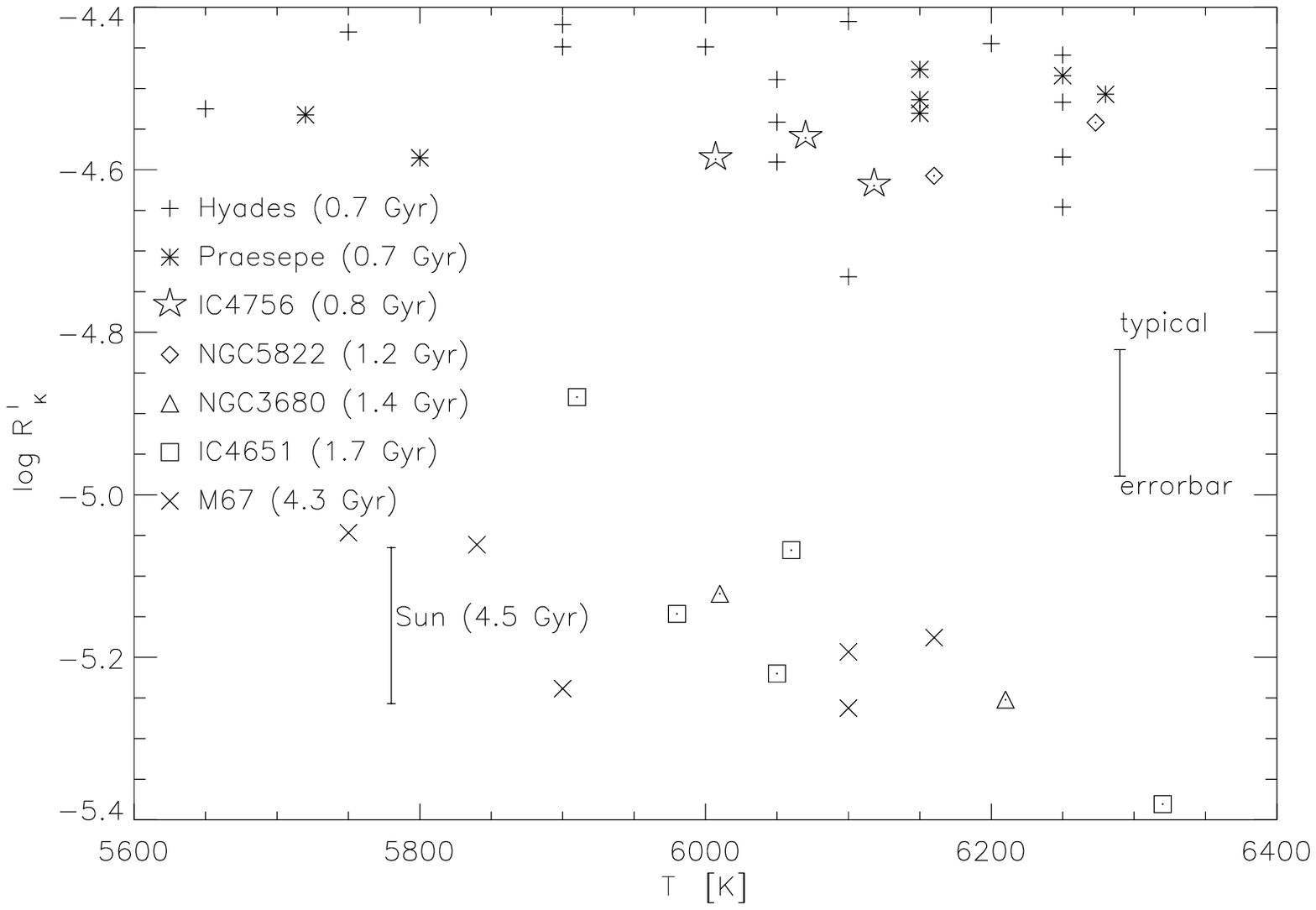}
&
\includegraphics[width=8cm]{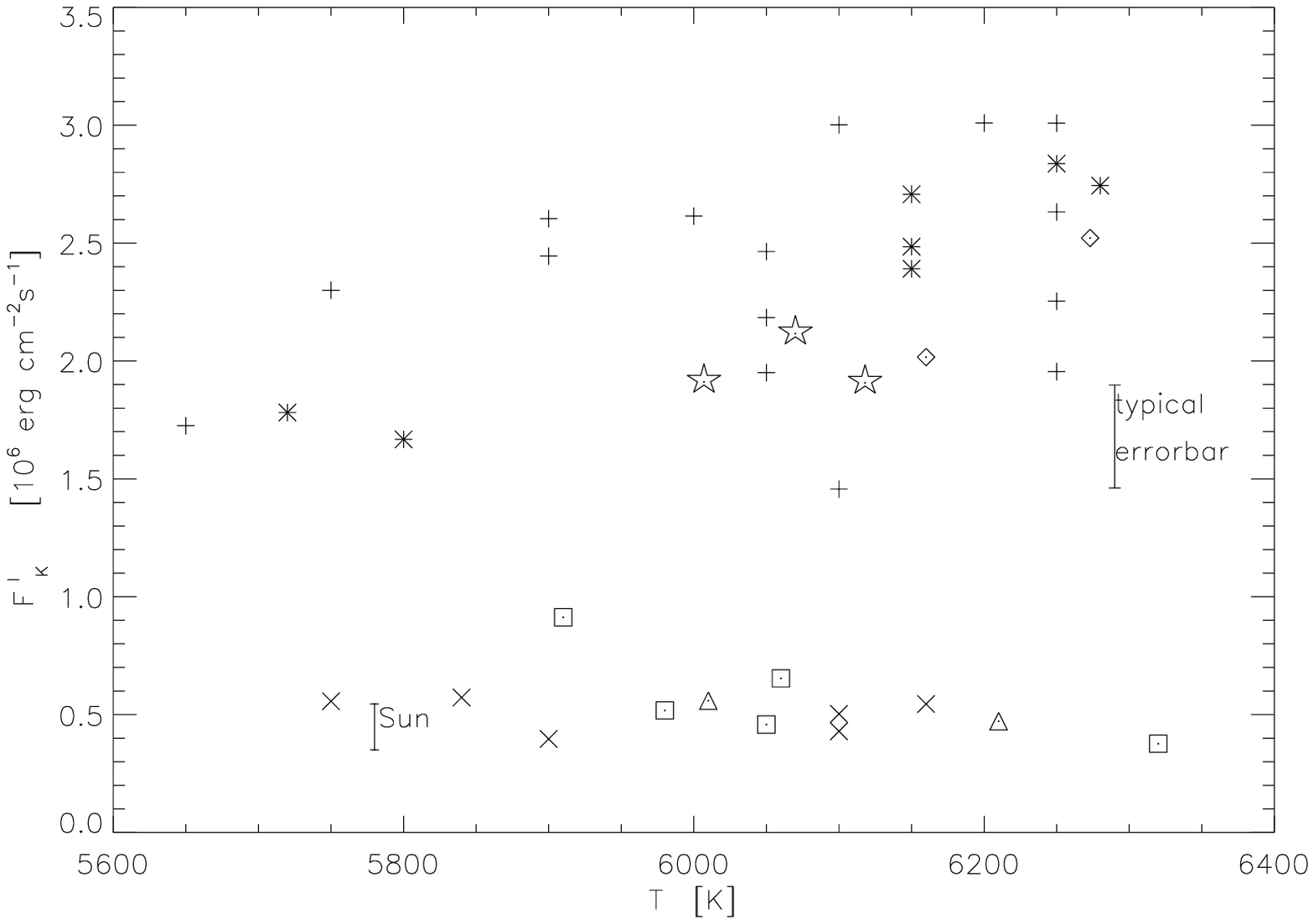}\\
\end{tabular}

\caption{Spectroscopic  temperature   versus  chromospheric  activity,
  using as a proxy  of chromospheric activity both $\log R^{\prime}_K$
  (left panel) and $F^{\prime}_K$ (right  panel). For the Sun, the bar
  corresponding to its  full variation during a cycle  is plotted. The
  typical  error bar shows  the typical  uncertainty ($\pm  1 \sigma$)
  evaluated from  the error analysis,  assuming an uncertainty  in the
  temperature of 110  K.  Had we just added the  counts to measure the
  1--{\AA}~K index in the new sample stars, which gives a secure lower
  limit  (see Sect.  \ref{CAda}),  we  would have  measured  in the  2
  NGC~5822 stars a value of $\log R^\prime_{K}$ lower by only 0.09 and
  0.07. The ages come from \cite{sal}. }

\label{fig} 
\end{center}
\end{figure*}

In Fig.  \ref{fig}  we plot the temperature of  the stars versus their
chromospheric  activity.   This  is  done  using as  an  indicator  of
chromospheric  activity  both  $\log  R^{\prime}_K$ (left  panel)  and
$F^{\prime}_K$  (right panel).   The  data are  those  shown in  Table
\ref{table1}.   The   distribution  of  chromospheric   activities  is
markedly  bimodal: the  more active  stars  are those  in the  Hyades,
Praesepe, IC~4756, and NGC~5822, while the inactive stars are those in
IC~4651, NGC~3680, and M~67 and the Sun.  Within either group there is
little  difference  between  stars  belonging to  different  clusters,
although  any small  correlation with  age using  averages  within the
active  group would  be obscured  by  the cyclic  range in  individual
stars.  But the two groups are separated by a wide range of activities
in which only two stars lie.   This range coincides with what is known
as the Vaughan--Preston gap, an underpopulated chromospheric--activity
range identified by \cite{vpgap} using  a large sample of stars in the
solar neighbourhood.

According to these ages and to  the data shown in Figure \ref{fig}, no
strong age--activity correlation appears  to exist after 0.7 Gyr, just
a rapid transition from active to inactive that occurs in a very short
time,  namely  200  Myr, between  the  age  of  NGC~5822 and  that  of
NGC~3680. With this data it seems unlikely that a strong age--activity
relationship could be masked by the  scantiness of the data, or by the
activity   variations.   Thus   we  can   confidently   conclude  that
chromospheric activity drops much  faster between the ages of NGC~5822
and NGC~3680  than it does  in the  half Gyr before  and in the  4 Gyr
after. We would expect most of the  stars in the gap to be young stars
in an inactive phase or old stars in an active phase rather than stars
at a certain age.  In our  sample we have one in either category.  Two
alternative explanations might be  considered.  Either our two targets
in  NGC~5822  were both  in  an  exceptionally  active phase  when  we
observed them, or the age difference between the oldest active and the
youngest  inactive  cluster  is  much  larger  than  that  claimed  by
\cite{sal}.  In  the former  case, NGC~5822 solar--type  members would
normally lie in  the middle of the Vaughan--Preston  gap. Owing to the
extremely small sample of stars at  this crucial age, it is well worth
follow-up spectroscopic  observations of  the two NGC~5822  members to
accurately measure their time--averaged chromospheric activity, and to
add to the sample single members in the interesting temperature range.
The possibility  that the age  of NGC~5822 has been  over-estimated by
\cite{sal} might be seriously considered because it was not determined
directly, but by means of a photospheric index calibrated for a sample
of 11  clusters.  But  in this case  its age  would need to  have been
overestimated by almost a factor of two.

According to  \cite{papoc1}, \cite{paulson} and  the chemical analysis
of  the present  sample  of  stars, the  metallicity  of the  clusters
studied ranges  from [Fe/H] = -0.04  dex in NGC~3680 to  [Fe/H] = 0.27
dex in  Praesepe.  Since  the 3 coeval  active clusters   have similar
levels   of   chromospheric    activity   despite   having   different
metallicities ([Fe/H]=0.01,  0.13 and 0.27 dex),  metallicity does not
appear to play a major role within the limited range of values spanned
by our clusters.

Chromospheric   activity   undergoes   large   variations   on   short
time--scales.  For  instance, the  chromospheric activity of  the Sun,
from the highest  activity peak to the Maunder  minimum, spans a range
that  goes from  the lower  boundary  of the  Vaughan--Preston gap  to
slightly above the  level of the most inactive stars  (only 8\% of the
solar neighbourhood  stars less active  than the lowest  solar value).
This   range,   according   to   the  current   calibration   of   the
age--versus-$\log R^\prime_{HK}$  relationship, corresponds to  an age
interval from 2.5  to 8 Gyr \citep{henry}. However,  members of binary
or multiple systems have a similar level of chromospheric activity. It
is  mainly  this circumstance  that  led  \cite{sod91}  to claim  that
chromospheric  activity   correlates  well  with  age   and  allows  a
calibration. In  the light of  the results presented here,  we suggest
alternative explanations should be sought.

Another calibration of the  chromospheric activity evolution with time
is that  of \cite{lach99}. From  Fig. 4 therein,  it can be  seen that
their result agrees with ours  as far as inactive stars are concerned,
i.e.  chromospheric activity does not  evolve after it has crossed the
Vaughan--Preston gap, at least until  the solar age.  As far as active
stars  older  than  the  Hyades  are  concerned,  there  are  4  to  6
data--points (for two  stars it is not possible to  say whether or not
they  are younger  than  the Hyades  and  therefore out  of the  range
considered  in  the  present  study).   For this  group,  the  Pearson
coefficient indicates a level of anticorrelation between age and $\log
R^\prime_{HK}$ that is insignificant (-0.27) or fair (-0.57), depending
on  how  many stars  we  consider. We  conclude  that  these data  are
compatible with the  conclusion that, within the age  range from 0.7
to  1.2  Gyr  and  from  1.4  Gyr  to  solar  age,  any  age--activity
relationship must be weaker than the short term variations.

From \cite{MH} it  is clear that the one open  cluster that supports a
strict monotonicity of the chromospheric activity time evolution (once
short--time  scale  variations are  smoothed  out)  in  the age  range
between  that  of  the  Hyades  and  that of  M~67  and  the  Sun,  is
NGC~752. This cluster, according to \cite{sal}, is older than NGC~5822
and  younger than NGC~3680,  i.e.  exactly  in the  range in  which we
expect the transition to occur.   Therefore, again, these data are not
incompatible with  the scenario we suggest, in  which the intermediate
chromospheric activity level of NGC~752 would be due to its age.

The two groups  of stars on opposite sides  of the Vaughan-Preston gap
differ not only in their chromospheric activity level, but also in its
trend  with  the  colour/temperature.   This  can be  seen  in  Figure
\ref{fig}, and  it is much  more evident when a  larger spectral--type
range  is considered  \citep[see  e.g.][]{MH}.  Furthermore,  temporal
variations  of  chromospheric activity  are  large  and irregular  for
active  stars  and  small  and  regular for  inactive  ones.   Several
attempts have  been made to  provide a physical explanation  for this,
for example  \cite{dmr81} proposed  a transition from  a complex  to a
simpler magnetic--field  morphology which occurs at the  time when the
rotation decreases enough to  reach a threshold value.  More recently,
\cite{barnes}  detected two  sequences  in the  period--versus--colour
diagram of  open clusters, and  he associated them with  two different
rotation  morphologies,  intertwined  with  stellar  magnetic  fields.
\cite{bv07} suggested a change of dynamo mechanism to explain the fact
that  stars   occupy  two  very  distinct  sequences   in  a  rotation
period--versus--cycle period diagram. The  kind of scenario in which a
transition of the nature of the dynamo takes place at a specific point
of  the  stellar lifetime  could  be  used  to explain  the  phenomena
reported and discussed by us.

\begin{acknowledgements}

  It  is a  pleasure to  thank Boris  Dintrans for   long  and useful
  discussions on  the dynamo effect.  We also thank  the referee
    for  useful   comments.   Data  collected  at   ESO,  VLT.   This
  publication  made use  of  data products  from  the WEBDA  database,
  created by  J.-C- Mermilliod and  now operated at the  institute for
  Astronomy  of the  University  of Vienna.   The SIMBAD  astronomical
  database and  the NASA's  Astrophysics Data System  Abstract Service
  have also  been extensively used.   G.  P., J.   M.  and N.   C.  S.
  acknowledge  the  support  of projects  PTDC/CTE-AST/65971/2006  and
  PTDC-CTE/AST-66181/2006  of the  Portuguese Funda\c  c\~{a}o  para a
  Ci\^{e}ncia e a Tecnologia.

\end{acknowledgements}

\bibliographystyle{aa}

\Online

\begin{table}
\begin{center}
\begin{tabular}{c c c c c}
Star&T$_{eff}$   &1--{\AA}  K &$F^{\prime}_K$& $\log R^\prime_{K}$ \\
    &           &index       & [10$^6$ erg/ &                   \\
    &[K]        &      [m\AA]&  cm$^2$ sec]             & [dex] \\
\hline
\multicolumn{4}{c}{}\\                    
\multicolumn{4}{c}{Hyades, 0.7 Gyr}\\                    
 van Bueren   1 & 6250 &180 & 1.95 $\pm$ 0.17  &-4.65  $\pm$ 0.04\\
 van Bueren   2 & 6050 &223 & 2.18 $\pm$ 0.24  &-4.54  $\pm$ 0.05\\
 van Bueren  10 & 6100 &281 & 3.00 $\pm$ 0.31  &-4.42  $\pm$ 0.04\\
 van Bueren  15 & 5750 &308 & 2.30 $\pm$ 0.34  &-4.43  $\pm$ 0.06\\
 van Bueren  17 & 5650 &272 & 1.73 $\pm$ 0.29  &-4.53  $\pm$ 0.07\\
 van Bueren  18 & 5900 &294 & 2.60 $\pm$ 0.33  &-4.42  $\pm$ 0.05\\
 van Bueren  31 & 6200 &261 & 3.01 $\pm$ 0.28  &-4.44  $\pm$ 0.04\\
 van Bueren  49 & 6050 &207 & 1.95 $\pm$ 0.21  &-4.59  $\pm$ 0.05\\
 van Bueren  52 & 6050 &248 & 2.46 $\pm$ 0.27  &-4.49  $\pm$ 0.05\\
 van Bueren  65 & 6250 &202 & 2.25 $\pm$ 0.20  &-4.58  $\pm$ 0.04\\
 van Bueren  66 & 6250 &253 & 3.01 $\pm$ 0.27  &-4.46  $\pm$ 0.04\\
 van Bueren  73 & 6000 &270 & 2.61 $\pm$ 0.30  &-4.45  $\pm$ 0.05\\
 van Bueren  88 & 6250 &228 & 2.63 $\pm$ 0.23  &-4.52  $\pm$ 0.04\\
 van Bueren  97 & 5900 &279 & 2.44 $\pm$ 0.31  &-4.45  $\pm$ 0.05\\
 van Bueren 118 & 6100 &161 & 1.46 $\pm$ 0.15  &-4.73  $\pm$ 0.04\\
 \multicolumn{4}{c}{}\\                    
\multicolumn{4}{c}{Praesepe, 0.7 Gyr} \\                    
 KW  49         &6150 &233  & 2.39  $\pm$ 0.25 & -4.53  $\pm$ 0.04\\
 KW 100         &6150 &255  & 2.71  $\pm$ 0.28 & -4.48  $\pm$ 0.04\\
 KW 208         &6280 &238  & 2.74  $\pm$ 0.25 & -4.51  $\pm$ 0.04\\
 KW 326         &5800 &234  & 1.67  $\pm$ 0.25 & -4.59  $\pm$ 0.06\\
 KW 368         &5720 &270  & 1.78  $\pm$ 0.29 & -4.53  $\pm$ 0.07\\
 KW 392         &6250 &248  & 2.84  $\pm$ 0.26 & -4.48  $\pm$ 0.04\\
 KW 418         &6150 &239  & 2.48  $\pm$ 0.26 & -4.51  $\pm$ 0.04\\
\multicolumn{4}{c}{}\\   
\multicolumn{4}{c}{NGC 3680, 1.4 Gyr}\\        
                   
Eggen 60        &6010 & 95 $\pm$  5  & 0.56 $\pm$ 0.09& -5.12$\pm$ 0.07\\
Eggen 70        &6210 & 84 $\pm$ 10  & 0.47 $\pm$ 0.16& -5.25$\pm$ 0.14\\
\multicolumn{4}{c}{}\\                    
\multicolumn{4}{c}{IC 4651, 1.7 Gyr}\\  
                        
AMC 1109 &6060 &102$\pm$ 10 & 0.65 $\pm$ 0.14  &  -5.07 $\pm$ 0.10\\
AMC 2207 &6050 & 85$\pm$ 10 & 0.46 $\pm$ 0.13  &  -5.22 $\pm$ 0.13\\
AMC 4220 &5910 &133$\pm$ 20 & 0.91 $\pm$ 0.24  &  -4.88 $\pm$ 0.12\\
AMC 4226 &5980 & 93$\pm$ 15 & 0.52 $\pm$ 0.18  &  -5.15 $\pm$ 0.15\\
Eggen 45 &6320 & 74$\pm$ 10 & 0.38 $\pm$ 0.16  &  -5.38 $\pm$ 0.18\\
\multicolumn{4}{c}{} \\                    
\multicolumn{4}{c}{M 67, 4.3 Gyr} \\ 

Sanders  746 & 5750 &108 $\pm$ 20 & 0.56 $\pm$ 0.20 & -5.05 $\pm$ 0.16 \\
Sanders 1048 & 5900 & 85 $\pm$ 10 & 0.40 $\pm$ 0.12 & -5.24 $\pm$ 0.13 \\
Sanders 1092 & 6160 & 88 $\pm$ 15 & 0.54 $\pm$ 0.22 & -5.18 $\pm$ 0.17 \\
Sanders 1255 & 5840 &105 $\pm$ 20 & 0.57 $\pm$ 0.22 & -5.06 $\pm$ 0.17 \\
Sanders 1283 & 6100 & 82 $\pm$ 15 & 0.43 $\pm$ 0.21 & -5.26 $\pm$ 0.21 \\
Sanders 1287 & 6100 & 90 $\pm$ 10 & 0.50 $\pm$ 0.14 & -5.19 $\pm$ 0.12 \\
\multicolumn{4}{c}{}\\                    
\multicolumn{4}{c}{NGC 5822, 1.2 Gyr}   \\                           
  TATM 11003& 6160 &197$\pm$8&2.02$\pm$0.22&-4.61$\pm$0.05\\
 TATM 11014& 6270 &217$\pm$7&2.52$\pm$0.24&-4.54$\pm$0.04\\
\multicolumn{4}{c}{}\\      {}\\                     
\multicolumn{4}{c}{IC 4756, 0.8 Gyr}   \\                            
    HER 165& 6070 &211$\pm$7&2.12$\pm$0.23&-4.56$\pm$0.05\\
    HER 240& 6010 &206$\pm$6&1.91$\pm$0.22&-4.59$\pm$0.05\\
     HER 97& 6120 &190$\pm$9&1.91$\pm$0.22&-4.62$\pm$0.05\\
\multicolumn{4}{c}{}\\                    
\multicolumn{4}{c}{solar cycle 21, from \cite{wl81} }   \\       
Sun &5780&95$\pm 10$& 0.45$\pm$0.10\\

\end{tabular}

\caption{The Sample with the 3 Different Chromospheric Indices.}

\label{table1}

\end{center}
\end{table}

\end{document}